\documentstyle[preprint,aps]{revtex}

\begin{document}

\draft    


\title{Pairing in cuprates from high energy 
       electronic states}
       
\author{A.F. Santander-Syro$^{1}$, R.P.S.M. Lobo$^{1}$, 
        N. Bontemps$^{1}$, \\
        Z. Konstantinovic$^{2}$, Z.Z. Li$^{2}$ and H. Raffy$^{2}$}

\address{$^{1}$Laboratoire de Physique du Solide, CNRS UPR 5,
         Ecole Sup\'erieure de Physique et Chimie 
         Industrielles de la Ville de Paris,  
         75231 Paris Cedex 5, France}

\address{$^{2}$Laboratoire de Physique des Solides, CNRS UMR 8502,
         Universit\'e Paris-Sud 91405 Orsay cedex, France}       
         
\date{\today}

\maketitle

\begin{abstract}
The {\it in-plane} optical conductivity of 
Bi$_{2}$Sr$_{2}$CaCu$_{2}$O$_{8+\delta }$ thin films with 
small carrier density (underdoped) up to large carrier density 
(overdoped) is analyzed with unprecedented accuracy. Integrating 
the conductivity up to increasingly higher energies points to 
the energy scale involved when the superfluid condensate builds 
up. In the underdoped 
sample,  states extending up to 2~eV contribute to the superfluid. 
This anomalously large energy scale may be assigned to a change 
of {\it in-plane} kinetic energy at the superconducting transition, 
and is compatible with an electronic pairing mechanism.\\
\end{abstract}


In conventional superconductors, electrons bind into Cooper pairs by 
exchanging a phonon.  The condensation of 
pairs leads to the zero-resistance superconducting state. In cuprate 
superconductors, the binding mechanism remains an open question. 
One key issue is  the typical energy scale of the excitations  
responsible for pairing. 
Infrared (IR) and visible spectroscopy measures the charge density distribution as 
a function of energy, through the investigation of the area under the 
frequency ($\omega$) and temperature ($T$) dependent optical conductivity 
$\sigma _{1}\left( \omega ,T\right)$. 
This area, known as the spectral weight $W$, is defined as:
 	
\begin{equation}
 W = \int_{0^{+}}^{\Omega_{M}}\sigma _{1}\left( \omega ,T\right)  d\omega 
 \label{spectral weight}
\end{equation}
where $\Omega_{M}$ is a cut-off frequency. When integrating from zero to 
infinite frequency, this spectral weight should be conserved as it depends 
only on the total charge density and the bare electronic mass. 
Ferrell, Glover and Tinkham (FGT)
 noted that, in the 
superconducting state,  the spectral weight  $\Delta W$ lost  from
the finite frequency (regular) conductivity is retrieved in 
the spectral 
weight $W_s$ of the $\delta (\omega)$ function centered at zero frequency, 
representing the condensate \cite{FGT}.
Actually, $\Delta W$ is approximately equal to $W_s$ (the so-called FGT sum 
rule) as soon as the cut-off 
frequency $\Omega_{M}$ in Eq.1 covers the spectrum of excitations responsible for the 
pairing mechanism. In conventional superconductors, this occurs 
at an energy   corresponding roughly to $ 4 \Delta$ ($\Delta$ is the 
superconducting gap, related in BCS theory to the Debye frequency) \cite{FGT}. 
Assuming a similar behaviour in cuprates, the FGT rule should be exhausted
at $\hbar \Omega_M \sim  0.1$~eV, as a typical maximum gap value 
in these d-wave superconductors 
is roughly 25 meV \cite{GapHTc}. A violation of this sum rule, 
i.e. $\Delta W < W_s$ when integrating up to 0.1 eV  , 
was reported for the interlayer optical conductivity of some cuprate 
superconductors \cite{Basov,KatzBasov,BasovVDM}, and discussed as
possibly related to a change of interlayer kinetic energy. 
This question has indeed raised 
 active experimental and theoretical discussions 
\cite{Tsvetkov,Hirsch3,Sudip1,Sudip2}, connected in particular with the interlayer
tunneling theory \cite{Anderson}.

To date there is no experiment showing such unconventional behaviour 
directly for the in-plane conductivity of cuprates. Nevertheless this point
was also given a renewed interest \cite{Hirsch1,Hirsch2}. 
What is at stake is 
that the need of an energy scale higher than any typical phonon energy  
to exhaust the FGT rule would be the hallmark of an electronic mediated 
pairing mechanism. Until now, the
changes observed below the critical temperature $T_c$ in the 
in-plane optical response at large energy scales
remained inconclusive 
\cite{FugolVIS,HolcombVIS,VdMTrieste}. 
 
This paper demonstrates, from a thorough 
study of the FGT sum rule of the in-plane conductivity, that indeed an 
electronic energy scale is involved when underdoped 
Bi$_{2}$Sr$_{2}$CaCu$_{2}$O$_{8+\delta }$  (Bi-2212) becomes superconducting.

 Three thin films 
from the Bi-2212 family were selected, at three doping levels which 
probe three typical locations in the phase diagram: the underdoped (UD), 
the optimally doped (OPD) and the  overdoped (OD) regime.
 We find that retrieving the condensate 
spectral weight in the OD and OPD samples requires integrating up to 
an energy of the order of 0.1 eV (800 cm$^{-1}$), i.e. a conventional 
energy scale. In the UD sample, however, the integration must be performed up 
to $\sim 16000 ~{\rm cm}^{-1}$ (2~eV), an energy scale much larger than 
typical boson energies in a solid. 
The scenario which emerges hence favors an electronic pairing mechanism 
at low doping level.

The three films were epitaxially grown by r.f. magnetron sputtering on
(100) SrTiO$_3$ substrates. The maximum critical temperature (defined 
at zero resistance) obtained in these conditions is $\sim 84$~K. The OD 
and UD states were 
obtained by post-annealing the films in a controlled atmosphere \cite{Zorica}. 
X-Ray analyses confirmed that the films are single phase. Our 
films have the following characteristics: i) UD film: $T_c=70$~K,
thickness $\sim 2400$~\AA; ii) nearly--optimally doped (OPD) film:
$T_c=80$~K, thickness $\sim 4400$~\AA; iii) OD film: $T_c=63$~K, 
thickness $\sim 3000$~\AA. Their optical homogeneity was verified by 
infrared microscopy with a lateral resolution of $20 ~ \mu$m \cite{Asantan}. 
The reflectivities, taken at 15 temperatures between $300$~K and $10$~K,
were measured in the spectral range $\left[30 - 7000\right]~{\rm cm}^{-1}$
with a Fourier Transform spectrometer, supplemented with standard  visible
spectroscopy in the range $\left[4000 - 25000\right]~{\rm cm}^{-1}$. 
Using thin films rather than single crystals allows to measure reliably 
relative variations in reflectivity 
within less than 0.2~\%, due to their large surface (typically 6 x 6~mm$^2$).  

It is known that temperature changes of the optical response in the 
mid-infrared 
and the visible ranges are small, but cannot be neglected \cite{Maksimov}. 
Yet, most studies rely  on a single 
spectrum at one temperature in the visible range \cite{VdMTrieste}. We did 
monitor the temperature evolution of the reflectivity 
spectra in the full available range. This is obviously important if one 
is looking for a spectral weight transfer originating from 
(or going to) any part of the {\it whole} frequency range.

The contribution of the substrate to the experimentally measured
reflectivities precludes the Kramers-Kronig analysis in thin films. 
For all raw spectra, an accurate fitting 
(within $\leq 0.5$~\%), taking into account the substrate response 
at each measured temperature and the interference pattern in the film, 
was performed (Fig.1, top panel) \cite{Asantan,Bilayer}. This procedure
determines the dielectric function for Bi-2212, hence the optical conductivity. 
Moreover, the fit yields a valuable extrapolation of the 
conductivity in the low energy range ($\omega < 30 ~{\rm cm}^{-1}$,
not available experimentally) \cite{Quij}, which is
important in the evaluation of the spectral weight.

The conductivities at $T_{A}  \geq T_c$ and  $T_{B} < T_c$, 
are shown in Fig.~1 (lower panel), up to
800 cm$^{-1}$ (0.1~eV)
for the UD ($T_{A} = 80~K$) and OD ($T_{A} = 70~K$) samples respectively. 
 In both cases, $T_{B} = 10~K$. In the OD sample, the curve 
$\sigma _1\left(\omega,T \leq T_c \right)$ lies {\it below}
the one at $T \geq T_c$, exhibiting an
expected loss of spectral weight in this energy range. In contrast,  the curve 
$\sigma _1\left( \omega, T \leq T_{c} \right)$ for the UD sample lies {\it above} 
the one at $T \geq T_c$, up to 100~cm$^{-1}$, then crosses it, and no loss of
spectral weight is apparent in the energy range shown.

From an experimental point of view, the FGT sum rule usually compares 
the change $\Delta W = W(T_{A}) -W(T_{B})$ (Eq.1)  
 and the superfluid spectral weight $W_s$. 
 $W_s$ was determined for $T < T_c$ at low frequencies within the measured 
 spectral range, by looking at the region where the real part of the
 dielectric function $\varepsilon_1\left(\omega\right)$ 
behaves linearly when
plotted versus $1/\omega^2$ (London approximation). An example is shown in the inset 
of Fig.1. The slope is directly related 
the superfluid spectral weight $W_s$  through the ``London" frequency 
$\Omega_{L} = c / \lambda_L$, where $\lambda_L$ is the London penetration depth. At 10~K, 
for instance, we find $\Omega_{L} = 7200 ~{\rm cm}^{-1}$ and 
$2350 ~{\rm cm}^{-1}$ for the OD and UD samples respectively \cite{slope}. 

Figure 2 shows the ratio $\Delta W / W_s$  for the samples studied in 
this work. Note that the figure extends actually over three different energy scales. 
For the OD and OPD
samples, the sum rule is exhausted at roughly
$500-1000~{\rm cm}^{-1}$, i.e., 2.5 to 5 times the maximum gap, as in 
conventional superconductors. In the UD sample though, $\Delta W/W _s$ starts negative and 
becomes positive at $\sim 600 ~{\rm cm}^{-1}$. 
Even at energies as large as $8000~{\rm cm}^{-1}$, $\Delta W/W_s \sim 0.7$. 
It keeps increasing with increasing energy and approaches 1 at 
$\sim 16000~{\rm cm}^{-1}$. 
 A large part ($\sim 30$\%) of the
superfluid weight in the underdoped regime thus  builds up at the 
expense of spectral weight coming from high energy regions of the optical 
spectrum ($\hbar \omega \geq 1 \rm eV$).
 
This remarkable behavior must be critically examined in light of 
 the uncertainties that enter in the determination of the ratio $\Delta W/W_s$. 
Firstly, the determination of $\Delta W$ assumes that $W(T_{A})$ is a 
fair estimate of the normal 
state regular spectral weight $W_{n}(T_{B})$, if one could drive the
system normal below $T_c$. While this assumption is correct in BCS superconductors, 
it is no longer valid for High-$T_c$ superconductors \cite{Sudip1}, hence our 
taking the normal-state 
spectral weight $W(T_{A} \geq T_{c})$  instead of  $W_{n}(T_{B})$ 
(unknown) may bias the sum rule. The error incurred
by doing so can be estimated as follows.
Figure 3 displays the temperature dependence, from 300~K down 
to 10~K, of the relative spectral weight 
$W(\Omega_{M},T) / W(\Omega_{M},300$~K), for three selected integration ranges, 
according to Eq.1.  
 At $\Omega_{M}$=1000~cm$^{-1}$,  the normalized spectral weight  
 exhibits a significant increase as the temperature is lowered, 
and could therefore
keep changing in the superconducting state. One could infer an increase of 
$\sim 10$\% of this relative 
spectral weight for both samples if assuming a linear extrapolation 
of the data. Hence $W(T_{A})$ is most likely to give too small an estimate for 
$W_{n}(T_{B})$ at $T < T_c$, 
in this energy range. To get a better insight of this possible underestimate,
the superfluid weight $W_s(T)$ was 
added to the spectral weight $W (T_B)$ (open symbols in Fig.~3), at a 
frequency $\omega =\Omega _{M}$\ and a temperature $T < T_c$. The points 
extrapolate reasonably well the high temperature data, thus suggesting an upper 
bound for the error  \cite{UpBoundErrFGT}. These estimates have been performed 
for a number of cut-off frequencies starting from 100 cm$^{-1}$ (the error 
increases at low frequency).
At 5000~cm$^{-1}$ and above, relying on similar considerations about the 
temperature evolution of the relative spectral weights shown in Fig.3, 
the changes with temperature of the normal state spectral weight should be
approximately 10 times smaller than at 1000 cm$^{-1}$. 
Note that at 20000~cm$^{-1}$, the spectral 
weight is constant,
meaning that the relevant energy scale for the FGT sum rule has been achieved.
 Therefore, above 5000~cm$^{-1}$,  the uncertainty 
 in $W_{n}(T_{B})$ becomes negligible compared to those due to the error 
in the relative change with temperature (the error in the absolute value
being irrelevant), in the fitting,
 and in the determination of $W_s$.
 
The three latter uncertainties were calculated self-consistently 
since the error in the determination of the superfluid density is related to
the error in the determination of the optical conductivity. 
They yield an upper bound of $15~\%-20~\%$ 
in the uncertainty on the evaluation of $\Delta W / W_s$, for all frequencies.
All uncertainties  can then be 
incorporated in the analysis of the ratio $\Delta W / W_s$, and are 
represented by the error bars in Fig.~3. Therefore, the top of the error 
bars delineates the upper limit for the FGT sum rule at each frequency.

For the UD sample, it is then clear 
that the negative value could be assigned to an incorrect estimate of $W_{n}$. 
However, the violation of the sum rule for this sample, with 
$\Delta W/W_{s}=0.65 \pm 0.18$ at $8000 ~{\rm cm}^{-1}$ is also clearly 
established. Within the error bars, the sum rule is exhausted  
in this sample above $16000 ~{\rm cm}^{-1}$.
Underdoped Y-123 showed a more conventional behavior, possibly because,  
as suggested, only one spectrum is usually recorded in the visible range
which is precisely the energy range which matters in this case \cite{VdMTrieste}.
Our results for the OD and OPD samples agree with earlier similar work 
in YBa$_{2}$Cu$_{3}$O$_{7-\delta }$ (Y-123) and 
Tl$_{2}$Ba$_{2}$CuO$_{6+\delta }$ (Tl-2212)  \cite{BasovVDM}.

In the context of the tight-binding Hubbard
model, a relation exists between the low-frequency spectral
weight and the kinetic energy $E_{kin}$ per copper site \cite{VdMTrieste}: 

\begin{equation}
  \frac{\Delta W}{W_{s}}-\frac{4\pi c}{137\hbar}\frac{a^{2}}{V} %
  \frac{1}{\Omega_{L}^{2}} \left(E_{kin,s}-E_{kin,n}\right) = 1 
 \label{FGTEk}
\end{equation}
where $a$ is the (average) lattice 
spacing in the plane, $V$ is the volume per 
site (SI units). This relation 
means that a breakdown of the FGT sum-rule up to an energy 
$\hbar \Omega_M$ of the order of the plasma frequency ($\sim 1$~eV for Bi-2212) 
is related to a change in the 
carrier kinetic energy $\Delta E_{k} = E_{kin,s}-E_{kin,n}$, 
when entering the superconducting 
state. According to our results in the UD sample (Fig.2), 
$\Delta W/W_s=0.65 \pm 0.18$ at 1 eV, which yields
$\Delta E_{k} =1.1 \pm 0.3$~meV per copper site. This would be a 
huge kinetic energy gain, $\sim$ 15 times larger than the condensation 
energy  $U_0$. For (optimally) doped Bi-2212, 
$U_0 \simeq 1$~J$/$g-at~$\approx 0.08$~meV per copper site \cite{Loram}.
A change of the in-plane kinetic energy could actually drive the 
superconducting transition, as it has been proposed in various scenarios: 
holes moving in an antiferromagnetic background \cite{AFEk}, 
interlayer tunneling theory \cite{Anderson}, or hole undressing 
\cite{Hirsch1,Hirsch2}. The latter 
scenario suggests that the violation of the FGT sum rule
must be more conspicuous for a dilute concentration of carriers and that, 
upon doping, a conventional energy scale exhausting the FGT sum-rule 
should be retrieved. Also, the kinetic 
energy lowering $\Delta E_{k}$ may be much larger than the condensation energy,  
and was estimated for Tl-2212 to be $\sim 1-3$ meV 
par planar oxygen \cite{Hirsch3}, 
which results into $0.5-1.5$ meV per copper site.
It was also suggested that it should be easier to observe the sum 
rule violation in UD  samples in the dirty limit \cite{Hirsch3}, 
which could apply in our case. 

Recently, STM experiments in optimally doped Bi-2212 samples showed  
small scale spatial 
inhomegeneities, over $\simeq$ 14 \AA, which are reduced significantly 
when doping increases, and whose origin could be local variations of oxygen 
concentration \cite{Davis}. Since the wavelength in the full spectral range 
is larger than 14 \AA, the reflectivity performs a large scale average 
of such an inhomogeneous medium. The implications in the conductivity are 
still to be investigated in detail, but it is presently 
unclear how this could affect the sum rule.

In conclusion,
we have have found for the in-plane conductivity of the underdoped
Bi-2212 a clear violation of the sum rule at 1~eV,  corresponding to 
a kinetic energy 
lowering (within the framework of the tight-binding Hubbard model) of 
$\sim$ 1~meV per copper site.
The very large 
energy scale required in order to exhaust the sum rule in the UD sample 
cannot be related to a conventional bosonic
scale, hence strongly suggests an electronic pairing mechanism.\\

We are very grateful to M.~Norman and C.~P\'epin for illuminating discussions. 
We acknowledge
fruitful comments from J.~Hirsch,  and E.Ya.~Sherman. \ We thank P.~Dumas (MIRAGE beamline at LURE, Orsay) for his
help with the IR microscopy measurements. \ AFSS thanks Colciencias and
Minist\`{e}re Fran\c{c}ais des Affaires Etrang\`{e}res (through the Eiffel
Fellowships Program) for financial support. \ RPSML acknowledges the
financial support of CNRS-ESPCI.

\begin{figure}
\caption{Top panel: measured reflectivity spectra of the underdoped sample at 
         250~K and 10~K (open circles).
         Three peaks from the SrTiO$_{3}$
substrate are visible at 60, 180 and 530~cm$^{-1}$. 
The solid lines are the fitted spectra.
Bottom panel: real part of the
         calculated conductivity at temperatures above $T_c$ at 80~K 
         (open diamonds)
         and 70~K (open circles) and at 10~K (full symbols)
          for the UD (right scale) and OD
         (left scale) samples  respectively.
         The conductivities are extrapolated from 30~cm$^{-1}$ down to zero
         as a result of the fit.
         Inset: linear fit (solid line)of the low frequency 
         $\varepsilon_{1}\left(\omega,T<T_{c}\right)$ data from the 
         UD sample (open squares), versus $\omega^{-2}$.}
\label{fig1}
\end{figure}

\begin{figure}
\caption{Ratio $\Delta W / W_s$ versus frequency 
showing the exhaustion of the FGT sum rule at conventional energies for the 
OD (diamonds, right error bars) and OPD (triangles, middle error bars) samples.
 An unconventional ($\sim 16000~{\rm cm}^{-1}$ or 2 eV) energy scale is 
 required for the UD sample (circles, left error bars). 
  Note that the frequency scale changes at $800$ and $8000~{\rm cm}^{-1}$.
  }
\label{fig2}
\end{figure}
\begin{figure}
\caption{Effective spectral weight $W(T,\Omega_{M}) /W(300~K,\Omega_{M}$)
versus temperature for
         the underdoped (top panel) and overdoped (bottom panel) samples, at
         different cutoff frequencies $\Omega_M$ (full symbols). 
         $\Omega _{M}=1000$ cm$^{-1}$ (circles).
         5000~cm$^{-1}$ (up triangles), and 20000~cm$^{-1}$ 
         (down triangles).           Open symbols are obtained by adding the 
         superfluid weight $W_{s}(T)$
         to the spectral weight $W(T<T_c)$.
         }
\label{fig3}
\end{figure}


\begin{references}
\bibitem{Bilayer}  The best description of our data was
  obtained by assuming no backward reflection from the rear of the substrate
  (i.e., assuming an infinitely thick substrate). Causality ensures that the 
  overall dielectric function is unique (within the error bars).
\bibitem{slope}  We chose specifically this procedure for the determination of the 
superfluid density because it does not require any extrapolation [(for instance 
of $\sigma_{2}(\omega)$] and ensures that the proper analytical form of 
$\varepsilon_{1}(\omega)$ is followed within the experimental range.
\bibitem{UpBoundErrFGT}  In the case of the underdoped sample, and below
  5000~cm$^{-1}$, we extrapolated  the data from $T \gtrsim T_{c}$ in Fig.3,
  hence maximizing the error in $W_{n}(T_{B})$.
\end{references}
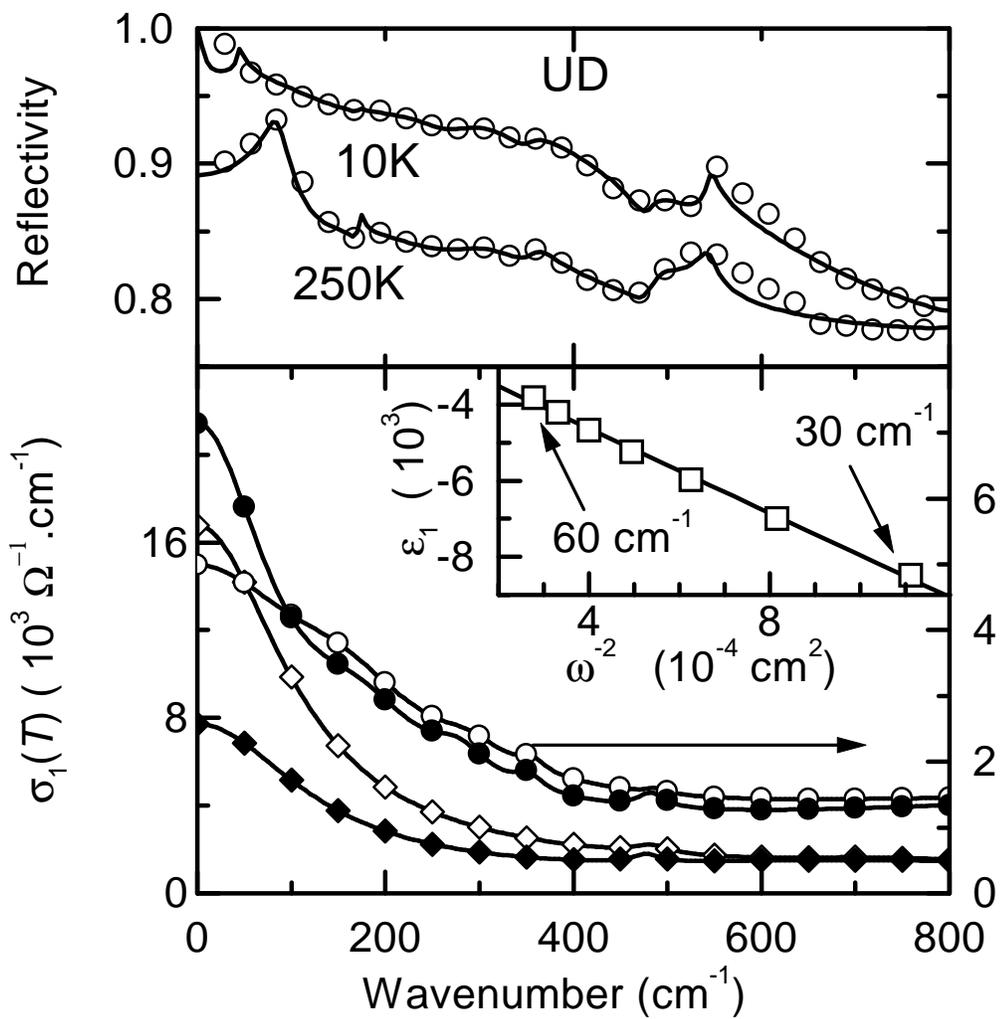
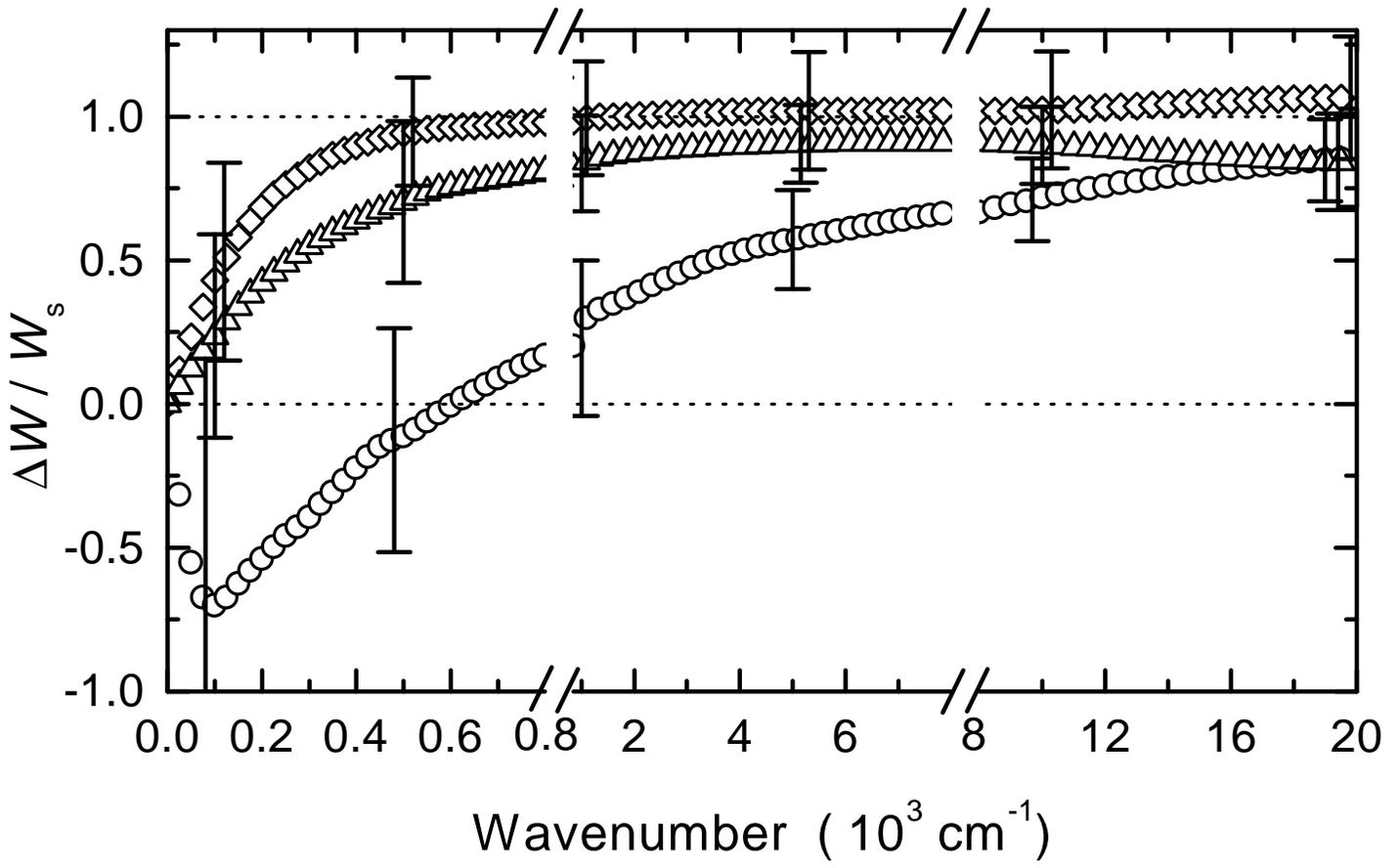
\end{document}